Article

# Ordinary Search Engine Users Carrying Out Complex Search Tasks

## Georg Singer and Ulrich Norbisrath
Institute of Computer Science, University of Tartu, Estonia

## Dirk Lewandowski
Department Information, Hamburg University of Applied Sciences, Germany

**Abstract**
Web search engines have become the dominant tools for finding information on the Internet. Due to their popularity, users apply them to a wide range of search needs, from simple look-ups to rather complex information tasks. This paper presents the results of a study that investigated the characteristics of these complex information needs in the context of Web search engines. The aim of the study was to find out more about (1) what makes complex search tasks distinct from simple tasks and whether it is possible to find simple measures for describing their complexity, (2) whether search success for a task can be predicted by means of unique measures, and (3) whether successful searchers show a different behaviour than unsuccessful searchers. The study includes 56 ordinary Web users who carried out a set of 12 search tasks using current commercial search engines. Their behaviour was logged with the Search-Logger tool. The results confirm that complex tasks show significantly different characteristics than simple tasks. Successful search behaviour can be distinguished from unsuccessful search behaviour. The implications of these findings for search engine vendors are discussed.

## 1. Introduction

For humans, information-seeking is a fundamental activity [1]. While seeking information was previously done in places such as libraries, nowadays, it is increasingly done through electronic media such as the Internet [2]. The ever-growing amount of information available on the Web increasingly overburdens Web users and impacts their Internet experience [3,4,5]. Search engines are the most prominent tools to help users find the information they are looking for. However, they do not support all information needs equally well. For example, complex Internet search is neither well-supported by current search engines nor well-researched [4], although there is strong agreement that search engines should support such tasks. According to research published by Microsoft [7], many queries issued to search engine "yield terrible satisfaction" and only 25% of the queries are successful. This might be due to those queries' being a part of search sessions. Those sessions can span longer time frames, with 50% of sessions lasting longer than one week.

Users often face a situation in which there is no simple answer to their information need. For example, a growing number of people use search engines to make decisions that require aggregated information [7]. Therefore, they have to review a large number of documents and/or synthesize results from different sources – discovering new factors that are relevant for their information needs along the way. A part of the problem is that we do not know enough about users' searching behaviour under these circumstances. While many studies on user behaviour have been conducted, studies seldom differentiate explicitly between simple and complex search tasks in the sense mentioned.

**Corresponding author:**
Georg Singer, Institute of Computer Science, Liivi 2, 50409, Tartu, Estonia
Georg.Singer@ut.ee

Today´s search systems are mainly designed to follow the "query–response" or short look-up concept. Search engines usually offer their users only a simple query interface. Users enter queries into those search systems and receive ranked lists of search results. Search engines support basic types of search tasks that can be solved with a simple query/result pairing. These search tasks usually happen in the context of question-answering and fact-finding [8]. We will call these *simple search tasks*. New studies show that the performance of current search engines for simple tasks is usually very good and satisfying for the users [9].

In this paper, we will investigate how users behave when dealing with *complex search tasks* (to be defined in the next section) and relate this to the support given by the search engines. Matching information needs to queries can be cumbersome for complex search tasks. Examples are open tasks such as comparing the development of various religions in different countries or finding the differences between the composers Mozart and Bach.

The tasks we study will remind the reader of exploratory search tasks. We see exploratory search tasks as being a subset of complex search tasks, as there are many tasks that require a lot of work (multiple queries, time-consuming, tabbed browsing) but do not necessarily carry all standard attributes of exploratory search tasks (such as learning or decision-making) [6,8]. For a more detailed discussion on classifying exploratory searches, compare with [10].

We conducted an experiment involving 56 users from mixed demographics carrying out 12 tasks of varying complexity. The users' activities were logged using the Search-Logger tool [11] (to be described in the Methods section). We analysed search behaviour in relation to the complexity of the tasks. In this paper, we want to suggest measures that will enable search engine providers to detect when a user performs a complex task and enable the providers to offer extended support. This quest for more support is also in line with research on Web Information Retrieval Support Systems (WIRSS) (e.g., [12]).

## 2. Related Work

We have divided our related work section into literature related to tasks, task complexity, logging user behaviour, and logging tools to cover all relevant aspects of the research presented in this paper.

### 2.1. Related work in tasks

"Tasks are activities people attempt to accomplish in order to keep their work or life moving on" ([l3], p. 1823). According to Vakkari [14], tasks usually have a goal, and information-searching is a contributing activity to find relevant information to achieve that goal. A work task is a special kind of task that appears in the work context, where the task's goal is also work-related. People usually carry out activities such as information-searching as the result of such work tasks or simply out of interest. Work tasks can trigger both information-seeking tasks and information-search tasks [15,16,17,18,19]. According to Li [l3], information-seeking tasks are defined as being related to people's general information needs. While seeking information, people typically search through multiple sources such as books in libraries, papers, and digital information systems. An information-seeking task usually becomes a search task once people start searching with IR systems

### 2.2. Related work in task complexity

Researchers disagree to a large extent on what makes a task complex [20]. There is a common understanding that task complexity can be either objective [21,22] or subjective [23]. The first definition is independent of the person carrying out the task. The second definition is dependent on the individual by looking at the cognitive demands [24] that are required of the performer of the task.

To find out more about what makes a task objectively complex, Li et al. [23] conducted a survey with 100 university students in China. According to their study, the main objective predictors for task complexity were the number of words in the task description, the number of domain areas that the task required, and the number of languages needed to interpret search results. In addition, the objective complexity criteria were more helpful in predicting complexity.

As far as subjective task complexity is concerned, it reflects how complex the person who carries out the task sees it [23]. It also relates to the number of sub-tasks that the user needs to carry out [25]. Byström and Järvelin [26] defined task complexity as information seekers' having to deal with "a priori determinability of, or uncertainty about, task outcomes, process, and information requirement" ([26], p. 194). According to Byström and Järvelin [26], search tasks are typically characterized by three types of information needs: problem information, domain information, and problem-solving information. The latter (i.e., known methods to tackle this problem) clearly distinguishes complex tasks from their simpler counterparts. Another classification of task complexity was developed by Campbell [20]. It is based mainly on a number of complexity-impacting factors that are present, such as multiple paths to a desired end-state, multiple desired end states, conflicting interdependence, and uncertainty or probabilistic linkages.

Singer, *et al.* [10] decomposed the complex search process into the aggregation, discovery, and synthesis steps. A complex search is defined as a multi-step and time-consuming process. It is not answerable with one query, needing instead synthesized information from more than one retrieved Web page or document. A complex search task is defined as one that leads to a complex search activity. Hence, a complex search task is described in relation to the complex search itself, which is an interactive process.

Finally, exploratory search tasks are defined as tasks related to open-ended information needs. They are usually abstract, not very well-defined, and often have a multifaceted character [6,8]. They are usually carried out requiring the searcher to apply high amounts of interaction while fulfilling needs such as learning, investigating, or making decisions. Another characteristic is that they are usually ambiguous and/or uncertain and require the searchers to discover new facets (problem aspects) of their search need along the way. It is important to state at this point that a number of search tasks require a lot of effort but still do not show all standard attributes of exploratory search tasks. A task might, for example, require a lot of interaction but no learning, planning, or decision-making. This kind of task is, therefore, complex but not necessarily exploratory, as defined before. Exploratory search tasks can be seen as a subset of complex search tasks.

## 2.3. Related work in logging user behaviour

Transaction logs recorded in search engines have been analysed in several publications [27,28,29]. One of the first larger-scale studies reflecting the search behaviour of Internet search users was conducted by Jansen, Spink, and Saracevic [30]. They analysed transaction logs from the search engine Excite. The authors reported that query modification was not happening very regularly, and the sessions were very short. Users seldom looked at more than two SERPs per session.

Jansen and Spink [29] gave an overview of nine transaction-log studies of five Web searches. They reviewed session length, query length, and query complexity and viewed content in different search engines. They observed that users were viewing very few results pages (even fewer than in their first study). Höchstötter and Koch [28] gave an extensive overview of query length and query complexity in Web searches and summarized their findings for multiple transaction log-based studies carried out between 1995 and 2008. They confirmed the findings of Jansen et al. [27]. They mentioned standard parameters to measure search engines and showed how these parameters reflected the online searching behaviour of search engine users.

Joachims et al. [31] carried out a study with 34 undergraduate students examining their interaction behaviour with SERPs – asking them to carry out 5 simple navigational and 5 informational search tasks. The study showed the preferences of users and their trust in search engines, as expressed by their clicking behaviour in the SERPs. Finally, research has confirmed that the majority of search sessions are short [27,28,29].

Most of the presented studies are only transaction-based and do not take the actual task and user into account. Other studies use a very specific user sample (such as undergraduate students). All of them confirm that queries are generally very short (only around two words, on average) and that users tend to look only at the first few results in the search engine results page (SERP). Most of the task-based studies were carried out with simple tasks. The only exception is the study described in Hölscher and Strube [32] that investigated more complex search tasks.

## 2.4. Related work in logging tools

Over the years, some logging tools have been developed to be used for research purposes, e.g. the "Wrapper" developed by Jansen et al. [33]. This tool was designed with a focus on evaluating exploratory search systems. Although its broad logging functionality is an advantage, the software does not offer the possibility of administering pre-composed search tasks to a group of users or relate user action to tasks and does not collect explicit user feedback. Another tool is a browser plug-in for Internet Explorer that was created by Fox et al. [34]. It was the first tool to simultaneously gather explicit as well as implicit information during searches. It evaluates the query level and gathers explicit feedback after each query. Fox's approach also came with a sophisticated analysis environment for the logged data. Unfortunately, Fox's Internet Explorer add-on is not publicly available. A third tool is Lemur's Query Log Toolbar [35], which is a toolkit implemented as a Firefox plug-in and an Internet Explorer plug-in. It logs implicit data on the query level. Other tools are the HCI browser [36], the Curious Browser [37], WebTracker [38], and Weblogger [39]. All the tools mentioned offer some kind of logging of user-triggered events. Still, none of these tools allows work task-based experiments (administering tasks to users and relating user action to tasks). As a consequence, we developed a logging tool called Search-Logger [11] to cost-efficiently carry out task-based user studies (for a more detailed description, please refer to the Methods section).

## 3. Research Questions

To guide our research, we formulated the following research questions.

**RQ1: Can we find simple measures to describe the complexity of search tasks?**
While it is clear that there are simple and complex tasks in Web searching, it is as yet unclear how to exactly distinguish between the two according to actual user behaviour. We analyse transaction logs to find relations between the monitored behaviour and the actual task that a user is carrying out.

**RQ2: What are measures for successful behaviour when carrying out a complex search task? What measures make successful searchers different from non-successful ones when carrying out complex search tasks?**
As the next step, we want to understand what makes successful search behaviour distinct. Therefore, we first look at the aggregated successful versus unsuccessful tasks, independent of the individual user. We aim to find measures for distinguishing a successfully completed complex search task from an unsuccessfully completed one. Then we compare good searchers (highly ranked over all search tasks in the experiment) with bad searchers (ranked low in the experiment) to rule out successful search behaviour based on pure luck.

## 4. Methods

### 4.1. Data Collection

The experiment was conducted in a laboratory environment in August 2011. The original user sample consisted of 60 volunteers who were being paid for their efforts (4 study participants had to be excluded from the experiment due to corrupt log data or insufficient computer skills). These volunteers were recruited using a demographic structure model to be representative for a cross-section of society in terms of age and gender (Table 1). The sample is relatively large for a lab-based user study, and with the wide span of users, the results of our study should not be limited to a certain user group such as university students (often taken in other studies). The experiment was conducted in Hamburg, Germany, and the language of the search tasks was German. The sample consisted of 32 women and 28 men, aged between 18 and 59.

**Table 1. Basic description of the user sample**

| Age Span | Female | Male | Total |
|---|---|---|---|
| 18-24 | 5 | 4 | 9 |
| 25-34 | 9 | 7 | 16 |
| 35-44 | 7 | 8 | 15 |
| 45-54 | 8 | 8 | 16 |
| 55-59 | 3 | 1 | 4 |
| Total | 32 | 28 | 60 |

## 4.2. Experimental design

The study participants were invited to the laboratory in four cohorts consisting of 15 people each. Before starting the experiment, all study participants were briefly instructed on how to use the Search-Logger tool (described below) for the experiment. After the instruction, each person was assigned a computer to be used throughout the experiment.

At each computer the Search-Logger tool [11] was installed. The Search-Logger is an add-on for Firefox browsers that allows administering work tasks to users. The Search-Logger runs in the background and does not interfere with the user's search behaviour. While the study participants carry out their assigned tasks, it automatically creates a log of a number of selected user events such as links clicked, queries entered, browser tabs opened and closed, bookmarks added and deleted, and automatically adds time stamps to each of those events. User specific information such as demographics and also user feedback about tasks is collected through automatic questionnaires at the beginning of the experiment and before and after each task. During the experiment, users can access their assigned tasks by clicking on the Search-Logger icon in the browser's status bar, which opens a new window. The tasks can be selected by using a drop-down button in this window. Selecting a task closes the Search-Logger window. During searches, users can switch between tasks – e.g. they can start with one task, pause it to work on another one, and return to the previously started task whenever they want. When they are satisfied with the amount of information they have found for a task, they can finish the task with a click. The users were assigned a set of 12 simple and complex search tasks (see below). They were given 3 hours to complete all the tasks.

## 4.3. Tasks

The search experiment consisted of 12 search tasks that can be classified as either simple or complex. Simple tasks typically allow users to find the required information in a single document. They can retrieve this document with a single right query. Complex tasks on the contrary are characterized by an open task description, accompanied by uncertainty and ambiguity and an open outcome [40]. Carrying out such a complex task typically requires issuing multiple queries, screening various Web sites, and discovering unknown aspects of a problem [10].

When creating the tasks, we made sure that the answer was available on German public websites as of August 2011; of course, the users were also allowed to search in other languages. We set up the sequence of tasks such that the users could alternatively solve simple and complex ones. The aim was to keep the participants interested and not to discourage them through a sequence of complex search tasks that they might be unable to solve. Users were allowed to switch between tasks (as described above). An overview of all the tasks is presented in Table 2.

**Table 2. Search tasks used for the experiment**

| Task Nr. | Task Type | Task Description |
|---|---|---|
| 1 | simple | When was the composer of the piece "The Magic Flute" born? |
| 2 | simple | How high was the state debt of Italy in comparison to their gross domestic product (GDP) in June 2011 in %? |
| 3 | simple | How many opera pieces did Verdi compose? |
| 4 | simple | When and by whom was penicillin discovered? |
| 5 | simple | Joseph Pulitzer (1847-1911) was a well-known journalist and publisher from the U.S. The Pulitzer Prize carries his name. In which European country was Pulitzer born? |
| 6 | simple | How many euros do you get if you exchange 10,000 units of the currency of Lithuania? |
| 7 | simple | How hot can it be, on average, in July in Aachen, Germany? |
| 8 | complex | What are the most important five points to consider if you want to plan a budget wedding? |
| 9 | complex | You were offered the job to run the local Goethe Institute (responsible for German language and cultural education) abroad. The chance is high that you will be sent to Astana |
| | | (Kazakhstan). Please collect facts and information (about half a page) about the political situation in Kazakhstan and the living quality. |
| 10 | complex | What is the name of the creature in the following picture and who is the author? Hint: this Austrian writer is also well-known in Germany. (Illustration omitted for copyright reasons) |
| 11 | complex | Are there differences regarding the distribution of religious affiliations between Austria, Germany, and Switzerland? What are they? |
| 12 | complex | There are five countries whose names are also carried by chemical elements. France has two (31. Ga – Gallium and 87. Fr – Francium), Germany has one (32. Ge – Germanium), Russia has one (44. Ru – Ruthenium), and Poland has one (84. Po – Polonium). Please name the fifth country. |

### 4.4. Logging user behaviour

Using the Search-Logger tool [11], we collected implicit user interaction data during the study by logging a number of standard events (see above).

### 4.5. Data analysis

Before analysing the data, we cleaned the sometimes messy data partly automatically and partly manually from log entries that originated from any kind of advertising or push services such as news or automatic bookmark updates. To analyse query reformulations, we went through all queries (independent of the search tool used) and analysed two queries in a row regardless of whether the second query was new (no word was the same), narrowing (contained more words), broadening (contained fewer words), or equal (same query). This approach might be simple compared to more elaborate approaches to query refinement [41] (see also the discussion of query reformulation approaches in Hearst [42], pp. 141-156), but it allows for a clear distinction and captures the most common query reformulation types in a distinctive way. In addition, we analysed equal queries regarding change of search tool (e.g., Web to image search or vice versa), spelling correction, or visiting earlier or number of SERPs visited.

The search results (which users submitted as Word documents) were analysed manually. Each task was compared with a sample solution and rated as follows:
1. Correct
2. Having correct elements
3. Not correct
4. User did not supply a solution

In the case of simple tasks, where by definition only one correct solution was available, a task was graded correct if it was the right result and complete. In the case of complex tasks, it was a bit trickier, as there was no single right solution possible. We graded a result as correct if it covered all the main aspects of the search needs. If the supplied solution covered only some aspects we graded it as "having correct elements". Results were graded as "not correct" if they contained the wrong information.

Of the 60 people who were scheduled to take part in the study, we had to exclude four people from our analysis. The reasons were Search-Logger malfunction (two study participants) and insufficient computer knowledge of the study participant (two users). We used standard qualitative and quantitative methods (such as 2 sample t-tests) to analyse the data, assuming a confidence interval of 95%.

## 5. Results

In this section, we present the results in relation to the research questions.

### 5.1. RQ1: Are there simple measures for the complexity of search tasks?

As we use the task complexity model by Singer et al. [10] that identified task search time as one important indicator of task complexity, we ordered the tasks according to the overall search time it took users to carry out the task (from low to high). Table 3 presents the mean values for selected complexity indicators. The first shows the mean values for those measures over the six tasks with the smallest search time and the six tasks with the highest search in the second row. As expected, all measures are significantly different between simple and complex tasks (as indicated by low p-values). The average task time over the complex tasks was almost four times higher than the average over the simple ones (427 sec. vs. 140 sec.). The second measure was the average number of sessions needed to complete the task. The complex search tasks were carried out using a larger number of sessions than in the case of the simple ones (1.1 vs. 1.0).

Next, we looked at the time that users spent on SERPs. This measure was approximately four times higher for complex search tasks than for simple ones: 122 sec. vs. 33 sec. Searchers spent more time on SERPs for complex search tasks than for simple search tasks.

The next measure was the time spent for scanning and reading results pages, depicted by "read time". This time was about three times as high for complex search tasks in comparison to simple tasks: 307 sec. vs. 107 sec. Searchers spent more time scanning and reading search results pages in complex search tasks.

**Table 3. Task measures for simple and complex tasks**

|  | Number of sessions | SERP time (sec) | Read time (sec) | Task time (sec) | Number of pages | Number of queries |
|---|---|---|---|---|---|---|
| Average overall 6 simple tasks (n=336) | 1.0±0.01 | 33±3 | 107±6 | 140±8 | 2.5±0.2 | 2.1±0.1 |
| Average over all 6 complex tasks (n=336) | 1.1±0.02 | 122±9 | 307±15 | 427±17 | 7.4±0.5 | 6.4±0.4 |
| p-value (unpaired *t* test) | <0.01 | <0.01 | <0.01 | <0.01 | <0.01 | <0.01 |

Next, we looked at the average number of pages visited when carrying out a task, which was more than twice as high as for the simple ones: 7.4 vs. 2.5. The number of queries issued during a task was three times as high for complex search tasks as for simple ones: 6.4 to 2.1. The average query length of the queries issued per task, as outlined in Table 4, was 60% higher for complex search tasks than for simple tasks: 4.4 vs. 3.1. Finally, we looked at how often queries were changed. We investigated five types of query changes: (1) how often queries were newly entered (the query was totally different from the one entered directly before), (2) how often queries were changed (at least one word of two consecutive queries was the same), (3) how often queries were narrowed (one or more search terms were added to the query), (4) how often a query was broadened (a term was deleted from a query), and (5) how often users visited more than the first SERP (i.e., the query stayed the same, but more results were requested).

The number of new queries was approximately twice as high for complex search tasks as for simple tasks: 2.0 vs. 1.2. In the case of changed queries, this number was seven times higher for complex search tasks than for simple ones: 1.4 vs. 0.2. In the case of narrowed queries, the number for complex tasks was about three times as high as for simple tasks: 0.7 vs. 0.2. For broadened queries, the number was about 10 times higher for complex search tasks than for simple ones: 0.34 vs. 0.03. The number of SERPs in addition to the first one was about three times higher for complex than for simple tasks: 0.3 vs. 0.1. The difference between the total number of queries and the sum of the number of specific query changes came from queries where users navigated back to the first SERP and where users used the search engine´s spelling correction function. Finally, the number of browser tabs opened during the task was higher for complex tasks than for simple ones (2.8 vs. 2.2).

**Table 4. Query-related task measures for complex and simple search tasks (continuation of Table 3)**

|  | Query-length (words) | New queries | Changed queries | Narrowed queries | Broadened queries | More than 1st SERP | Number of tabs |
|---|---|---|---|---|---|---|---|
| Average over all simple tasks (n=336) | 3.1±0.1 | 1.2±0.03 | 0.2±0.04 | 0.2±0.03 | 0.03±0.01 | 0.1±0.02 | 2.2±0.04 |
| Average over all complex tasks (n=336) | 4.4±0.2 | 2.0±0.1 | 1.4±0.1 | 0.7±0.07 | 0.3±0.04 | 0.3±0.06 | 2.8±0.11 |
| p-value (unpaired *t* test) | **<0.01** | **<0.01** | **<0.01** | **<0.01** | **<0.01** | **<0.01** | **<0.01** |

## 5.2. RQ2: What are measures for successful behaviour when carrying out a complex search task?

In this section, we focus on the question of what makes successful search behaviour unique and what measures we can use to identify this successful search behaviour and distinguish it from unsuccessful search behaviour. First, we compared all correctly carried out tasks with wrongly carried out tasks independent of which user carried out the task. Then we compared the tasks carried out by the best searchers in the experiment with the tasks carried out by the least successful searchers.

**Table 5. Comparison of correctly and wrongly carried out tasks (averages over tasks)**

|  | Number of Sessions | SERP time (sec) | Read time (sec) | Task time (sec) | Number of pages | Number of queries |
|---|---|---|---|---|---|---|
| Correctly carried out complex tasks (n=149) | 1.07±0.02 | 96.8±12 | 258±16 | 355±22 | 6.4±0.5 | 5.2±0.4 |
| Wrongly carried out complex tasks (n=52) | 1.08±0.04 | 97.1±19 | 257±31 | 354±41 | 5.4±0.7 | 4.4±0.6 |
| p-value | >0.1 | >0.1 | >0.1 | >0.1 | >0.1 | >0.1 |

Table 5 shows a comparison between 149 successfully carried out and 52 wrongly carried out complex tasks. None of the measures is significantly different between correctly and wrongly carried out tasks (indicated by high p-values). When going into more detail about how queries were reformulated (as outlined in Table 6), none of the measures except for number of browser tabs opened was significantly different. The number of tabs opened was significantly higher for correctly carried out tasks: 2.9 vs. 2.4.

**Table 6. Query-related measures for correctly and wrongly carried out complex tasks (continuation of Table 5)**

|  | Query length | New Queries | Changed queries | Narrowed queries | Broadened queries | More than 1st SERP | Number of tabs |
|---|---|---|---|---|---|---|---|
| Correctly carried out tasks (n=151) | 4.3±0.2 | 1.9±0.1 | 1.1±0.12 | 0.6±0.07 | 0.3±0.05 | 0.2±0.04 | 2.9±0.2 |
| Wrongly carried out tasks (n=52) | 4.9±0.5 | 1.5±0.1 | 0.8±0.18 | 0.3±0.11 | 0.2±0.07 | 0.5±0.19 | 2.4±0.1 |
| p-value (unpaired *t* test) | >0.1 | 0.10 | >0.1 | 0.05 | >0.1 | >0.1 | **0.017** |

Finally, we wanted to rule out the case where a searcher is excellent in one instance but performs badly on all other tasks (which would not make him a good searcher and his search behaviour would just randomly be good). In contrast to the above analysis, where we compared successfully carried out tasks with unsuccessful ones, independent of which user carried out the search, we next compared good searchers with bad searchers. We created a ranking of the users according to their performance in the whole experiment. We ranked the users first by the number of correct answers and then, in cases of users with the same number of correct answers, by answers with correct elements. Table 7 compares the mean values for various measures for the users in the first quartile of the ranking (14 users, ranked 1 to 7) with the means over all measures for the users in the fourth quartile of the ranking (14 users, ranked 18-25).

**Table 7. Task measures for first and fourth quartile of users ordered by their ranking in the experiment**

|  | Number of sessions | SERP time (sec) | Read time (sec) | Task time (sec) | Overall search Time (sec) | Number of pages | Number of queries |
|---|---|---|---|---|---|---|---|
| 1st quartile (n=14) | 1.09±0.03 | 82±13 | 256±26 | 337±31 | 2027±184 | 6.7±0.8 | 5.9±0.5 |
| 4th quartile (n=14) | 1.16±0.06 | 166±29 | 360±57 | 526±68 | 2969±411 | 8.4±1.5 | 7.6±1.6 |
| p-value | >0.1 | **0.02** | >0.1 | **0.02** | 0.05 | >0.1 | >0.1 |

Successful users spent less time on SERPs than unsuccessful ones: 82 sec. vs. 166 sec. The average time per task was significantly smaller for successful searchers than for unsuccessful ones: 337 sec. vs. 526 sec.

**Table 8.** Query-related task measures for the first and fourth quartile of users ordered by their ranking in the experiment (continuation of Table 7)

|  | Query Length | New queries | Changed queries | Broadened queries | Narrowed queries | More than 1st SERP | Number of tabs |
|---|---|---|---|---|---|---|---|
| 1st quartile (n=14) | 3.8±0.4 | 2.0±0.2 | 1.2±0.2 | 0.7±0.1 | 0.4±0.05 | 0.2±0.06 | 4.0±0.6 |
| 4th quartile (n=14) | 5.7±1.3 | 1.9±0.2 | 1.5±0.3 | 1.0±0.4 | 0.4±0.11 | 0.5±0.16 | 3.0±0.6 |
| p-value | >0.1 | >0.1 | >0.1 | >0.1 | >0.1 | >0.1 | >0.1 |

None of the query-related task measures outlined in Table 8 shows significant differences between the best and worst searchers.

## 6. Discussion

Complex search tasks can be distinguished from their simple counterparts using standard measures. Our findings show that, for complex search tasks, all time-based measurements (i.e., search time, time on SERPs, and reading time) are higher than for simple search tasks. Whilst this may sound trivial at first, our findings confirm that time-based measures may be a good indicator for search engines to offer help within a search task/session. Users seldom use help files offered by the search engines, but contextual help may be useful in supporting users in achieving their search goals. The larger average number of sessions in complex search tasks can be an indicator of users' having difficulty in completing the search tasks. In addition to the longer time needed for completing the tasks, users may not be successful in their first try to solve the search task. They may use the break (and the time they spend with other search tasks) to re-think the task. In Web searches, short queries are usually used. Users break down their information needs into just a few keywords and assume that a search engine will present results suitable for this information need. It is only seldom that users enter whole sentences or questions into the search boxes. Studies have shown that this behaviour is even rarer in German-language queries than in English-language queries [43]. We therefore assume that, when users enter long queries or questions, they may be unsure about how to express their information need. Therefore, long queries may be a good indicator for search engines to offer some support to the user. Also, query changes show that a user is not successful with his initial query, so also query reformulations may be a good indicator for offering support. All in all, the results of our study confirm our idea about the complexity of the given tasks. Simple user behaviour-based measures are suitable for determining the complexity of a search task. This information is useful, as a search engine can use these measures to determine what kind of support to offer to a user.

The second research question was whether there are measures that enable us to identify search behaviour indicating successful task performance. We did not get a clear result here. The only significant result is that successfully carried out tasks show a significantly higher number of browser tabs opened and closed during the task in comparison to unsuccessfully carried out tasks. All other measures show too-high p-values to be accepted as significant. As our user sample was comparably small (n=56) and the users had diverse search experience (ranging from an inexperienced housewife to an experienced student of information science), we assume that a larger sample in combination with a more homogeneous average search experience would lead to smaller standard errors and clearer results. However, the high standard errors of mean show that Web searching is very diverse and that care should be taken when extrapolating results from small user studies to a larger population.

For a search engine provider, it should theoretically be possible to project the positive or negative outcome of a search task in terms of success by using simple measures. We assume here, however, that the search engine is unable to identify an individual user but, instead, needs to identify task complexity just from the current search session (using information from just the current search task). Nevertheless, deriving more information from the individual search session is a problem that should be investigated further. Whilst it is surely a trend with search engines to collect as much information on individual users as possible, we see a high need for data avoidance and data minimization as a principle.

We also analysed whether searchers consistently showed successful search behaviour or whether their search successes were just random. Our results confirm that successful searchers spend less time on SERPs and have a smaller task time. The fact that successful searchers spend less time on the SERPs could also result from their experience in examining SERPs. As SERPs are increasingly 'crowded' with results from a wide variety of sources and the results presentation changes from a simple list (where each result is presented as equal) to an extended list presentation (universal search) [44], there is an increased need for orientation on the SERPs. We conclude that more experienced searchers are able to get an overview of the results more quickly than inexperienced searchers.

One remark on preparing work tasks and controlling work task complexity is that the results confirm that the choice of tasks met our expectations regarding task complexity considerably well. All complex tasks (as outlined in Table 2) triggered complex search behaviour (reflected, e.g., by longer search times, multiple queries, and tabbed browsing). Also, all simple tasks except for one task triggered simple search behaviour. Task (3), "Italy Public Debt", was the only task that we had assumed to be simple, but in terms of total search time, it showed characteristics of a complex search task. We were surprised that it took the users so much time (on average, 331 sec.), resulting mainly from a high reading time. We assume that many of the study participants might have struggled with the difficulty of the topic (economics), so it took them longer to qualify the results that they found.

## 7. Conclusion

Complex search tasks are neither well-researched nor well-supported by current commercial search engines. The aim of our experiment was to find out how ordinary search engine users behave when confronted with search tasks requiring them to split the task into several elements and to synthesise information from a variety of documents. Overall, we can conclude that complexity can be expressed by the effort (in terms of time, sessions, queries, and browser tabs) needed to carry out a search task. This can be shown and proven by means of the measures that we investigated.

In terms of supporting the users of commercial search engines better, we suggest that search engine operators put more emphasis on the fact that complex search tasks have significantly different characteristics than simple ones. These differences can be measured as shown in this paper and, depending on the character of the search task, search engines could offer different kinds of support to the searcher. It is conceivable that the search process is monitored, not on a query basis (as it is now) but on a task basis. When erratic user behaviour is identified (such as identical queries being repeated) struggling searchers could be identified and offered help. We suggest that a different, enhanced service should be offered to these struggling searchers. In addition, it is conceivable that the whole search process is logged and a list of used queries visible to the user so that entering identical queries will be easier to avoid.

Our study also has its limitations, of course, e.g., the sample used. A wide variety of users participated in our experiment. While it was our intention to use a sample of users of a wide age span, and men and women alike, we had to pay the price in the form of wide variances in the measures used. Therefore, we were often unable to strongly confirm many of our hypotheses due to too-high standard errors of the means. This leads us to the conclusion that, in further studies, we should either use an even bigger sample or restrict our research to a certain user group. The latter approach, however, would not allow for universally valid results. We think that a large drawback of most user studies is exactly that they use only small samples (n<20) that are restricted to a certain user group, usually students from the researchers' universities.

Another limitation is that users were given limited time to complete the search tasks. Some users asked to be given more time after the three hours reserved for the experiment. Therefore, it could well be that some users would have (successfully) completed the tasks if they were given more time. However, the time given in the laboratory setting could also be a restriction the other way round: as users were given money to participate in the experiment, they may have been more willing to work on the difficult tasks than they may have been if they were working on the same tasks at home, not as part of an experiment. It is well-known that users put more effort into completing search tasks in a lab setting [45], and this limitation is not unique to our study. Another minor limitation is that all tasks were administered in the exactly same order. Although users were free to choose the task, the predefined task sequence in the menu could still have resulted in learning effects. In future studies, each user should be given a different task order. Also, it would be helpful to conduct extensive pre-testing of tasks to avoid events such as that which occurred with the task for Italy (Task 3), where the complexity assumed by the researchers differed considerably from the complexity as confirmed by the experiment. In the future, we plan to further analyse the data regarding patterns in the sequence of queries that the users used and whether those patterns could be used to identify strategies and erratic and chaotic search behaviour. It would also be interesting to investigate the differences between complex search tasks more deeply, where the complexity comes from the effort to aggregate, discover, or synthesise. Finally, it would make sense to carry out a similar experiment with a significantly larger user sample and more homogeneous user groups (such as, e.g., blue collar workers only).

## Acknowledgements

This research was supported in part by the Estonian Information Technology Foundation (EITSA) and the Tiger University program, as well as by the European Union (Archimedes).